\documentclass[twocolumn,showpacs,preprintnumbers,amsmath,amssymb,prl,superscriptaddress]{revtex4}

\usepackage{graphicx}
\usepackage{dcolumn}
\usepackage{bm}
\usepackage{dcolumn}
\usepackage{color}
\usepackage{units}

\begin{document}
\newlength{\LL} \LL 1\linewidth

\title{Colossal Spin Hall Effect in Ultrathin Metallic Films}

\author{Christian Herschbach}
\email{cherschb@mpi-halle.mpg.de}
\affiliation{Max Planck Institute of Microstructure Physics, Weinberg 2, 06120 Halle, Germany}
\author{Dmitry V. Fedorov}
\affiliation{Max Planck Institute of Microstructure Physics, Weinberg 2, 06120 Halle, Germany}
\author{Martin Gradhand}
\affiliation{H.~H.~Wills Physics Laboratory, University of Bristol, Bristol BS8 1TL, United Kingdom}
 \author{Ingrid Mertig}
\affiliation{Institute of Physics, Martin Luther University Halle-Wittenberg, 06099 Halle, Germany}
\affiliation{Max Planck Institute of Microstructure Physics, Weinberg 2, 06120 Halle, Germany}


\begin{abstract}
We predict spin Hall angles up to 80\% for ultrathin noble metal films with substitutional Bi impurities.
The colossal spin Hall effect is caused by enhancement of the spin Hall conductivity in reduced sample
dimension and a strong reduction of the charge conductivity by resonant impurity scattering. These findings
can be exploited to create materials with high efficiency of charge to spin current conversion by strain engineering.
\end{abstract}

\maketitle
An efficient transformation of charge into spin current is the key issue for future applications of
the spin Hall effect (SHE)~\cite{Dyakonov71,Hirsch99} in spintronic devices. The figure of merit for this
phenomenon is described by the spin Hall angle (SHA). For a long time, its largest value was $11\%$ measured
in Au~\cite{Seki08}. Subsequently, several studies revealed systems that provide SHAs
of comparable magnitude, the so-called \emph{giant} SHE~\cite{Seki08}. Among them, a SHA of $-12\%$ to $-15\%$
was measured in highly resistive $\beta$-Ta~\cite{Liu12}. Even larger values of $30\%$ to $33\%$ were obtained by
an experimental study of $\beta$-W thin films~\cite{Pai12}. Both results have been qualitatively predicted
evaluating a tight-binding model~\cite{Tanaka08}. 
In addition, a SHA of $-24\%$ has been reported for thin
film Cu(Bi) alloys~\cite{Niimi12}. The large magnitude of the SHA for a copper crystal with Bi impurities
was predicted by first-principles calculations relying on the skew-scattering mechanism~\cite{Gradhand10}.
The existing results inspire both theorists and experimentalists to search for new routes to synthesize
materials with even larger SHAs.

In this Letter we show that the SHA caused by impurities in ultrathin metallic films
can reach values up to 80\%. We concentrate on the skew scattering as the responsible
mechanism for the SHE in dilute alloys~\cite{Lowitzer11,Niimi11}.
The results are based on \emph{ab initio} calculations developed originally for bulk crystals~\cite{Gradhand10}
and later extended to the case of two-dimensional (2D) systems~\cite{Herschbach12}.
The parameter we are going to optimize is the spin Hall angle
\begin{equation}\label{eq.:SHA}
\alpha = \frac{\sigma_{yx}^s}{\sigma_{xx}}
\end{equation}
defined as the ratio of the spin Hall conductivity (SHC) $\sigma_{yx}^s$ to the longitudinal charge
conductivity $\sigma_{xx}$. This ratio is dimensionless since both conductivities
are expressed in the same units~\cite{Gradhand10,Fedorov13,Herschbach12}. For the skew-scattering mechanism
both depend inversely on the impurity concentration which provides $\alpha$ concentration independent.
Consequently, this quantity is ideally suited for comparison with experiment.

\begin{figure}[h!]
\vspace*{0.3cm}
\includegraphics[width=0.43\LL]{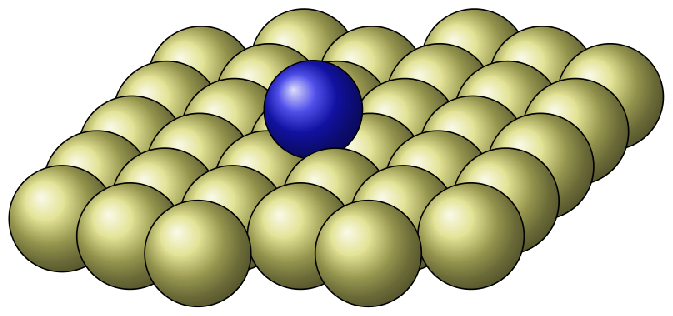}
\includegraphics[width=0.1\LL]{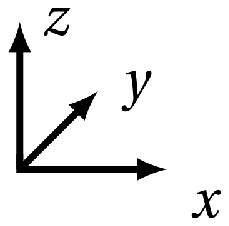}
\includegraphics[width=0.43\LL]{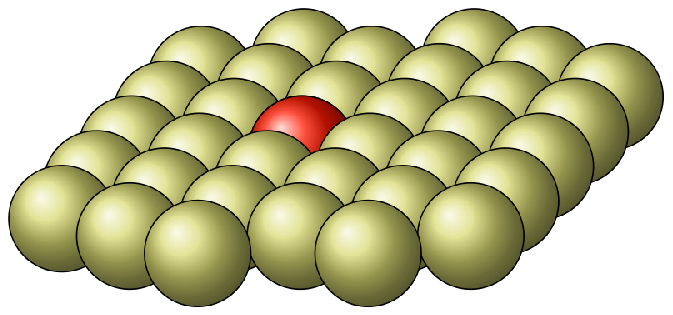}
\caption{(Color online) Geometry of the considered systems shown for a freestanding 1 ML fcc (111) film with
an adatom (left) or a substitutional impurity (right).}
\label{fig.:system_surf_ad}
\end{figure}

In our previous study~\cite{Herschbach12} we showed that the SHE induced by Pt impurities in Au(111)
films can be significantly enhanced for a reduced film thickness. Especially for the one monolayer (ML)
films a large SHA of $-13\%$ to $-18\%$ was obtained for the case of Pt adatoms~\cite{Herschbach12}.
For these systems the origin of the giant SHE was related to the enhanced effective spin-orbit coupling (SOC)
induced by the reduced coordination number and the associated strong potential gradients. In combination
with a single-sheeted Fermi surface it leads to the mentioned strong effects.
On the other hand, substitutional Bi impurities in bulk Cu provide a giant SHE as well. In this case
the large valence difference between impurity and host atoms causes strong scattering as well as strong
effective SOC. The idea of this paper is to combine both facts and screen the influence of Bi impurities
in ultrathin noble metal films. Accordingly, we focus on the 1~ML films considering Bi as adatom or
substitutional impurity. Figure~\ref{fig.:system_surf_ad} shows the general setup for our calculations.

\begin{figure}[h!]
\vspace*{0.3cm}
\includegraphics[angle=-90,width=0.92\LL]{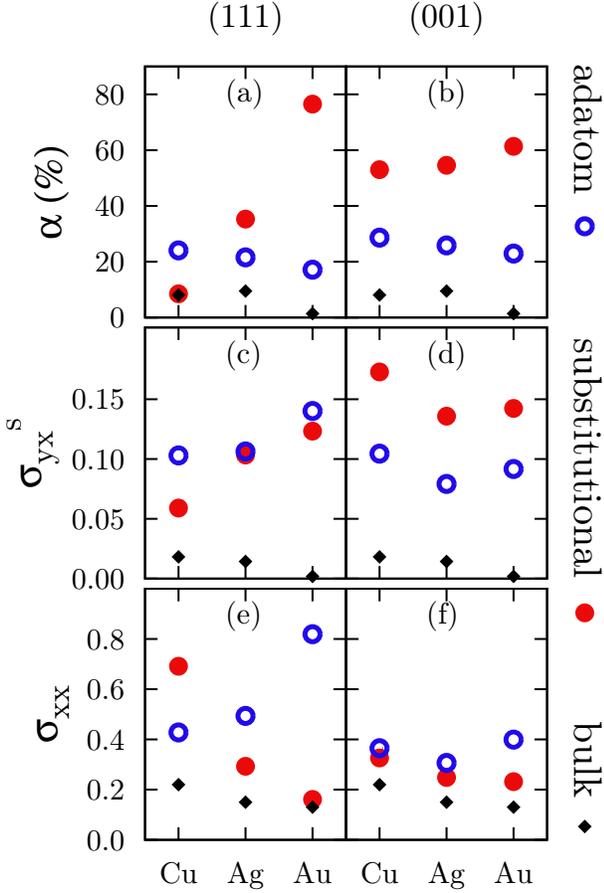}
\caption{(Color online) The spin Hall angle $\alpha$,
the spin Hall conductivity $\sigma_{yx}^s$,
and the charge conductivity $\sigma_{xx}$ for 1 ML (111) and (001) Cu, Ag, and Au films
with Bi as adatoms (blue rings) and substitutional impurities (red filled disks). For comparison,
the corresponding bulk values~\cite{Gradhand11} are shown with black diamonds. The conductivities
are given in units of $(\mu\Omega cm)^{-1}$ while $\alpha$ is shown in percent.}
\label{fig.:Alpha_sigma_1ML}
\end{figure}

The corresponding results obtained by the approach of Ref.~\onlinecite{Herschbach12} are presented
in Fig.~\ref{fig.:Alpha_sigma_1ML} where the SHA, the SHC, and the charge conductivity are shown for 1~ML (111)
and (001) noble metal films. Throughout the paper, the impurity density chosen for the films corresponds
to 1~at.\% impurity concentration used for the bulk crystals. First, nearly all values of the SHA shown
in Fig.~\ref{fig.:Alpha_sigma_1ML} (a)-(b) are drastically increased in comparison to the related bulk
systems with $\alpha < 10\%$~\cite{Gradhand11}. The SHAs for the (111) films are even larger than
the values reported in Ref.~\onlinecite{Herschbach12} for the corresponding hosts with Pt impurities ($\alpha < 19\%$).
Thus, combining the reduced coordination number of the impurity with the large valence difference between
impurity and host atoms seems to be a proper way to enlarge the SHA. In detail, the enhancement of the SHA
in comparison to the bulk results is solely determined by the SHC which is strongly increased in all systems,
as shown in Fig.~\ref{fig.:Alpha_sigma_1ML} (c)-(d). By contrast, the charge conductivity alone would cause
a reduction of $\alpha$, since it is also increased in comparison to the bulk values but enters Eq.~(\ref{eq.:SHA})
in the denominator.

Another striking feature is the strong dependence of the SHA on the host material for substitutional
Bi impurities in (111) films, culminating in the $80\%$ value for Au.
In contrast to the other configurations, $\sigma_{yx}^s$ and $\sigma_{xx}$ change for this case
in opposite direction going from Cu via Ag to Au, see Fig.~\ref{fig.:Alpha_sigma_1ML} (c) and (e). As a result,
both conductivities facilitate to amplify the corresponding SHA
shown in Fig.~\ref{fig.:Alpha_sigma_1ML}(a). However, the influence of the charge conductivity is stronger.
Within the (111) films
$\sigma_{xx}$ is reduced by a factor of 4 changing the host from Cu to Au. The particularly small charge conductivity
of Au(111) films with substitutional Bi impurities finally causes the colossal effect.
Therefore, the following detailed analysis will focus on substitutional impurities.

\begin{figure}[t]
\includegraphics[width=0.98\LL]{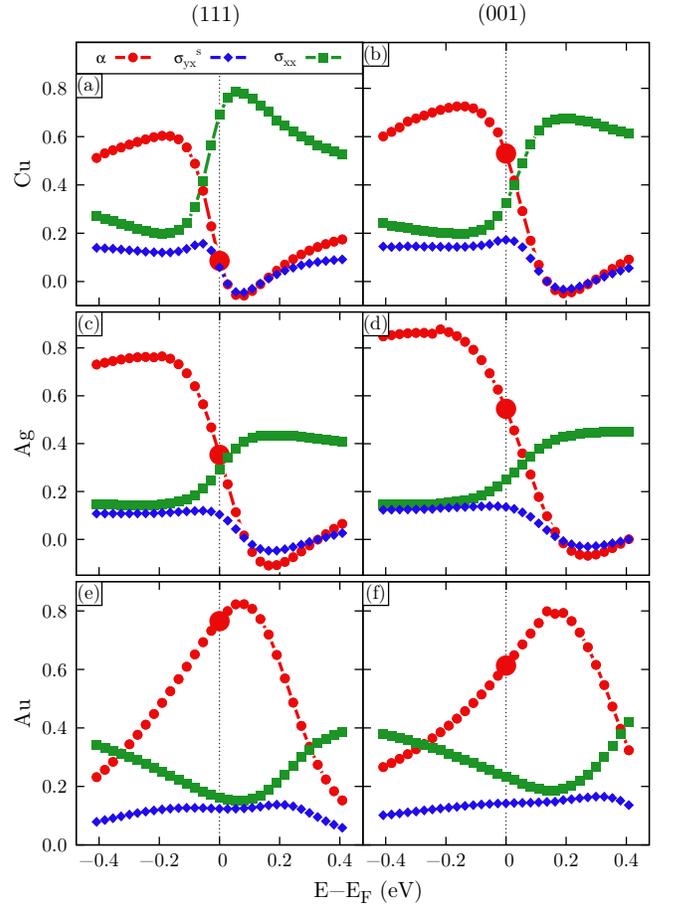}
\caption{(Color online) The results of transport calculations performed at different energies around the Fermi
level ($E_{F}$) are shown for 1 ML (111) and (001) Cu, Ag, and Au films with substitutional Bi impurities.
The corresponding $\alpha$ (red filled disks), $\sigma_{yx}^s$ (blue diamonds), and $\sigma_{xx}$ (green squares)
are displayed and $\alpha(E_{F})$ is highlighted with a larger disk. All quantities have absolute values in
the same order of magnitude but different units: the conductivities are in $(\mu\Omega cm)^{-1}$ while $\alpha$
is dimensionless and has to be multiplied by $100$ to get the values in percent.}
\label{fig.:DOS_tkkrdiffEF}
\end{figure}

In order to get a deeper insight
into the underlying mechanism of the colossal SHE, we investigate the energy dependence of the obtained quantities.
The corresponding values for the SHA and its constituents, $\sigma_{yx}^s$ and $\sigma_{xx}$, are shown
in Fig.~\ref{fig.:DOS_tkkrdiffEF}. In all cases $\alpha$ is strongly energy dependent. Here, it is important
to mention that the considered energy interval is so small that the topology of the single-sheeted 2D Fermi surface
of the 1 ML films~\cite{Herschbach12} remains unchanged.

Analyzing the contributions to the SHA from the charge and spin Hall conductivity separately, one can see
differences depending on the host material. While for Au the energy dependence of $\alpha$ is almost
solely determined by the charge conductivity, $\sigma_{yx}^s$ and $\sigma_{xx}$ play a comparable role for Cu
and Ag. Nevertheless, a certain correlation between the two conductivities exists for all
considered systems. Namely, the SHC increases with decreasing charge conductivity and vice versa,
which is most obvious for the Ag films. This is related to the fact that, generally, stronger
scattering should reduce $\sigma_{xx}$ but enhance $\sigma_{yx}^s$~\cite{Herschbach12}.
However, the situation in real systems can differ from such a simplified picture. For instance, the pronounced
maximum of the SHA in the Au films is merely determined by the corresponding minimum of the charge conductivity.
For both (111) and (001) Au films small changes of energy strongly affect $\sigma_{xx}$ while $\sigma_{yx}^s$
remains almost constant. In the case of the 1 ML Au(111) film the minimum of the charge conductivity occurs
in the vicinity of the Fermi level $E_F$, which causes the colossal SHA shown in Fig.~\ref{fig.:Alpha_sigma_1ML}(a).
Thus, going from bulk Au to the 1 ML Au(111) film, the giant SHE occurs since the SHC is increased due to enhancement
of the SOC and the absence of interband scattering~\cite{Herschbach12}. 
Finally, the SHA is further enhanced through the suppression of the charge conductivity.
These are the ingredients of the reported colossal SHA.

\begin{figure}[t]
\includegraphics[width=0.98\LL]{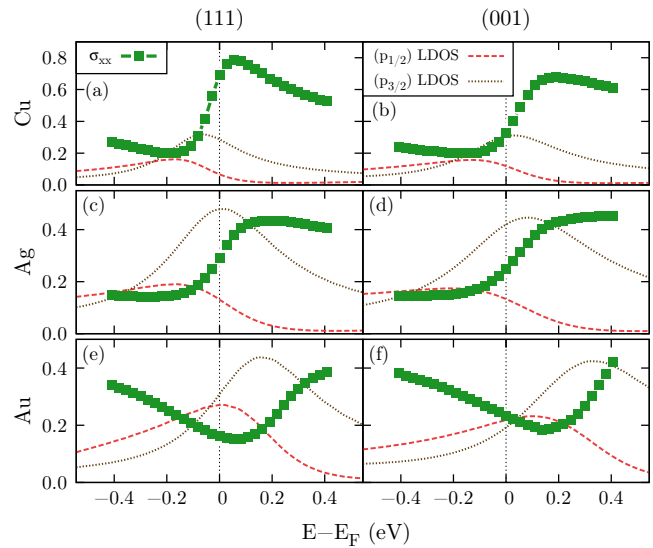}
\caption{(Color online) The contributions of the relativistic
$p_{\nicefrac12}$ channel (dashed red) and $p_{\nicefrac32}$ channel (dotted brown) to the impurity
local density of states (LDOS) are shown in comparison to the energy-dependent charge conductivity
$\sigma_{xx}$ (green squares) for 1 ML (111) and (001) Cu, Ag, and Au films with substitutional Bi impurities.
The charge conductivity and the impurity LDOS are plotted in units of $(\mu\Omega cm)^{-1}$ and $states$/eV,
respectively.}
\label{fig.:sigmaxx_noSOC}
\end{figure}

The origin of the reduced charge conductivity is resonant scattering, as demonstrated in Fig.~\ref{fig.:sigmaxx_noSOC}
presenting the impurity local density of states (LDOS). Bi impurities are known to be strong
$p$-scatterers~\cite{Niimi12,Gradhand10,Fedorov13}. Because of SOC the $p$-level is split
into $p_{\nicefrac 12}$ and $p_{\nicefrac32}$. The minimum of the charge conductivity is obviously correlated
with the maximum of the $p_{\nicefrac 12}$ LDOS of the Bi impurity (Fig.~\ref{fig.:sigmaxx_noSOC}).

Although the correlation between the $p_{\nicefrac12}$ impurity LDOS and the charge conductivity is obvious,
the underlying mechanism is quite complex. As shown in the Supplemental Material~\cite{Supplementary}, for all
considered systems the vertex corrections are of utmost importance for the energy dependence of $\sigma_{xx}$.
Being also the source of the considered skew-scattering contribution to the SHC, they occur in the used Boltzmann
equation by the so-called \emph{scattering-in} term~\cite{Gradhand10,Butler85}. This has a very subtle integral
structure and is obtained via an iterative procedure within our first-principles calculations~\cite{Supplementary}.
Therefore, it is difficult to provide a simple explanation for the relation between the impurity LDOS and the charge
conductivity.

Nevertheless, the strong influence of the vertex corrections is clearly connected to the resonance scattering.
Indeed, the well-pronounced peaks of the impurity LDOS near $E_F$, shown in Fig.~\ref{fig.:sigmaxx_noSOC}, are
missing in the case of substitutional Bi atoms in bulk Au~\cite{Supplementary} and
Cu~\cite{Tauber12}. As a consequence, the energy dependence of the charge conductivity
for bulk crystals is practically unaffected by the scattering-in term~\cite{Supplementary}.

With this microscopic picture in mind, it is possible to understand the host dependence of $\alpha$
for the two surface orientations. As illustrated by Fig.~\ref{fig.:DOS_tkkrdiffEF},
the general energy dependence $\alpha (E)$ is extremely similar between (001)
and (111) films for each host material. However, in the (001) case the reduced
coordination number causes a smaller charge density. As a result, the related
impurity resonance is shifted to higher energies with respect to the Fermi level. The conductivity
minimum is shifted accordingly. In addition, strong changes of $\sigma_{xx}$ at energies around
$E_F$ cause strong variations of $\alpha$. Thus, it is somewhat accidental that
the SHA shown in Fig.~\ref{fig.:Alpha_sigma_1ML} is nearly the same for the (001) films while
it varies strongly for the (111) films. Shifting the Fermi level of the (001) films slightly to
higher energies would simulate the situation of the (111) surface orientation.

\begin{figure}[t]
\includegraphics[angle=-90,width=0.95\LL]{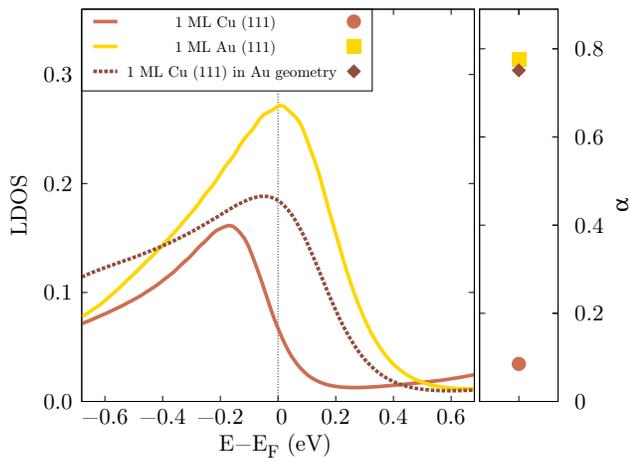}
\caption{(Color online) Left: The
$p_{\nicefrac12}$ local density of states at a substitutional Bi impurity atom in 1 ML fcc (111) film of Cu
(solid coppery, dark), Au (solid golden, bright), and Cu  with the gold lattice constant (dashed brown).
Right: The corresponding SHA calculated at $E_{F}$ for Bi in Cu (circle), Au (square),
and Cu with the gold lattice constant (diamond).} 
\label{fig.:CuinAugeo}
\end{figure}

This finding can be employed to optimize the SHA. The only condition required is to fix
the $p_{\nicefrac 12}$ impurity resonance at the Fermi level. A possible opportunity is
strain engineering of the film grown on an appropriate substrate. To prove this assumption, let us
choose the Cu(111) film showing the smallest SHA among all the considered systems, as indicated
by Fig.~\ref{fig.:Alpha_sigma_1ML} (a)-(b). According to Fig.~\ref{fig.:sigmaxx_noSOC}, the peak
of the corresponding $p_{\nicefrac 12}$ impurity LDOS is below $E_{F}$. Following the discussion
above, we can assume that an increase of the host lattice constant should provide a shift of
the impurity resonance towards the Fermi level. Figure 5 shows that the desired condition can be
fulfilled if the lattice constant of Au is used for the hypothetically strained Cu(111) film.
The increase of the lattice constant by about 13\% from $a_{\rm Cu}=3.615$~\AA~to $a_{\rm Au}=4.078$~\AA\ 
shifts the maximum of the $p_{\nicefrac12}$ impurity LDOS close to $E_F$. This in
return leads to the colossal SHA as shown in the right part of Fig.~\ref{fig.:CuinAugeo}. Of course,
the considered change of the lattice constant is quite strong. Nevertheless, the presented results
indicate a new route to design materials with large $\alpha$.

A substrate is needed in general, since the freestanding 1 ML films considered in our study are quite artificial
from a practical point of view.  To have more realistic systems for observation of the colossal SHE, a possible
way is to grow these films on an insulating substrate. However, in such a case the SHE can be modified by
the Rashba-type SOC~\cite{Wang13}, forcing electron spins to be oriented within the film plane. By contrast,
the considered SHE is largest for spins pointing out of plane. In order to ensure the desired
spin orientation, a symmetric quantum well structure {\it insulator}/{\it metal film}/{\it insulator} could be
utilized.

In summary, we predict a colossal spin Hall effect related to spin Hall angles up to 80\%, which are obtained
for ultrathin noble metal films with substitutional Bi impurities. This strong effect occurs for systems where
the giant SHE, caused by the reduced dimension of samples, is further amplified by a minimal charge conductivity
caused by resonant scattering. This condition is achieved if the impurity $p_{\nicefrac12}$ resonance state is
pinned at the Fermi level. The required resonant scattering can be tuned by strain engineering, depositing
films on an appropriate substrate. Our findings offer a new route to design materials with a very efficient
conversion of charge current into spin current.

This work was partially supported by the Deutsche Forschungsgemeinschaft (DFG) via SFB 762. In addition, M.G. acknowledges
financial support from the DFG via a research fellowship (GR3838/1-1). We also acknowledge the technical support of Katarina
Tauber with respect to the calculation of the energy-dependent conductivities.


\end{document}